\documentclass[useAMS,usenatbib]{mn2e}
\usepackage[dvips]{graphicx}
\usepackage{url}
\usepackage{amsmath}

\title[Highly reddened type Ia supernova SN 2004ab]{Highly reddened type Ia supernova SN 2004ab: another case of anomalous extinction}

\author[N. K. Chakradhari et al.]
{N. K. Chakradhari$^1$\thanks{Centre for Mega Projects in Multiwavelength Astronomy, Pt. Ravishankar Shukla University Raipur}, D. K. Sahu$^2$, G. C. Anupama$^2$, T.P. Prabhu$^2$\\
1. School of Studies in Physics \& Astrophysics, Pt. Ravishankar Shukla University, Raipur 492010, India\\
2. Indian Institute of Astrophysics, Koramangala, Bangalore 560 034, India\\
E-mail : nkchakradhari@gmail.com (NKC), dks@iiap.res.in (DKS), gca@iiap.res.in (GCA), tpp@iiap.res.in (TPP)}

\begin{document}

\date{Accepted .....; Received ......}

\pagerange{\pageref{firstpage}--\pageref{lastpage}} \pubyear{2017}

\maketitle

\label{firstpage}

\begin{abstract}
We present optical photometric and spectroscopic results of  supernova SN 2004ab, a highly reddened normal type Ia supernova.  The  total reddening is estimated as  $E(B-V)$ = 1.70 $\pm$ 0.05 mag. The intrinsic decline rate parameter, $\Delta m_{15}(B)_\text{true}$ is  1.27 $\pm$ 0.05, and $B$-band absolute magnitude at maximum  $M_{B}^{\text{max}}$  is $-$19.31 $\pm$ 0.25 mag. The  host galaxy NGC 5054 is found to exhibit anomalous extinction with very low value of $R_V$ = 1.41 $\pm$ 0.06 in the direction of SN 2004ab.
Peak bolometric luminosity is derived as $\log L_\text{bol}^\text{max}$ = 43.10 $\pm$ 0.07 erg\,s$^{-1}$. The photospheric velocity  measured from absorption minimum of Si\,{\sc ii} $\lambda$6355 line  shows  a velocity gradient of $\dot{v}$ = 90 km\,s$^{-1}$\,d$^{-1}$,  indicating that SN 2004ab is a member of the high velocity gradient (HVG) subgroup. The ratio of strength of Si\,{\sc ii} $\lambda$5972 and $\lambda$6355 absorption lines, $\cal R$(Si\,{\sc ii}) is estimated as 0.37, while their pseudo equivalent widths suggest that SN 2004ab belongs to  broad line (BL) type subgroup.

\end{abstract}


\section{Introduction}
\label{sec04ab_intro}
The correlation of absolute magnitude of type Ia supernovae (SNe Ia) with their observed properties such as the  decline in $B$-band magnitude  from its peak to 15 days after peak 
\citep{phil93,phil99}, stretch parameter \citep{perl97}, shape of the light curve \citep*{ries96}, colour \citep{rein05,wang06} and spectroscopic parameters \citep{nuge95,bene05} have made SNe Ia very important astronomical events, 
as they provide means to calibrate the luminosity at maximum.
Further, the high luminosities of SNe Ia enable us to see them at very far distances in the universe and use them as a standard candle for distance estimation.  

The observed properties and theoretical investigations suggest that SNe Ia are a result of the explosion of a Carbon-Oxygen white dwarf (WD, \citealt{Hoyl60}) that gains mass via accretion in a binary system or due to merging with another WD. The binary companion may be a main sequence/red giant star (single-degenerate scenario; \citealt{whel73}) or another WD (double-degenerate scenario; \citealt{iben84,webb84}). As the mass of the WD reaches close to Chandrasekhar limit \citep{chan31},  an instability sets in,  leading to thermonuclear runaway fusion reaction disrupting the WD. However, the exact nature of the progenitor and explosion scenario are still debated \citep*{maoz14} and need to be addressed properly for  use of SNe Ia in precision cosmology \citep{howe11}.

An estimate of the reddening suffered by SNe Ia and to correct for it   is very important for using them as distance indicators. Though majority of SN host galaxies show extinction properties similar to that of Milky Way,  consistent with total to selective extinction ratio of $R_V$ = 3.1,  this value is found to be significantly lower in the direction of SNe Ia in many host galaxies \citep{kris06,elia06,kris07,wang08,fola10,aman14}. Departure of $R_V$ from that  of Galactic value is generally referred as non-standard extinction.  Several studies   suggest that non-standard extinction with $R_V$ lower than 3.1 is shown mostly by SNe Ia which are significantly reddened (\citealt*{jhas07}; \citealt{fola10,chot11}; \citealt{mand11}; \citealt{scol14}). 

The reddening of SNe Ia arises from at least two different sources: the first one, causing small amount of reddening seen in most SNe Ia is consistent with the properties of interstellar dust in the Milky Way, while the second source responsible for reddening in the highly-extincted objects is characterised by an unusually low value of $R_V$ \citep{phil12}. \citet{wang05} and \citet{goob08} suggest that low value of $R_V$ could result from multiple scattering of light due to dust in the  circumstellar medium (CSM).  
It is shown by \citet{goob08} and \citet{aman14} that extinction properties of highly-reddened objects with low value of $R_V$ follow  a power law.

\begin{figure}
\centering   
\resizebox{\hsize}{!}{\includegraphics{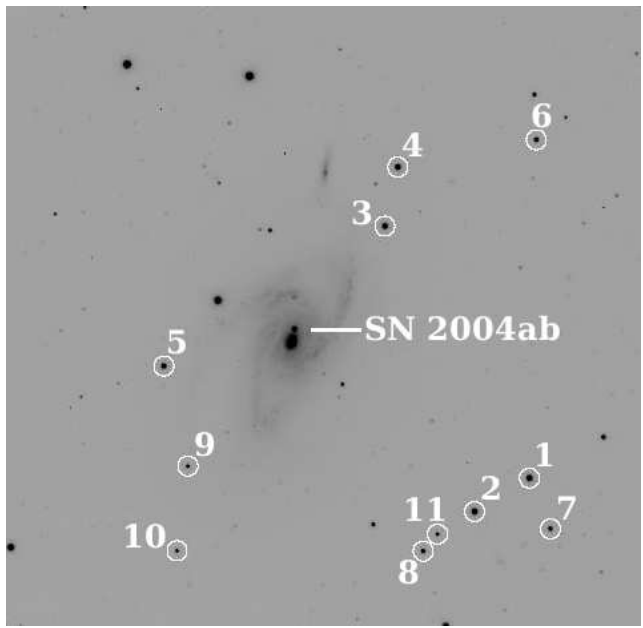}} 
\caption[]{Identification chart for SN 2004ab. The stars used as local secondary standards are marked as numbers 1--11 and their calibrated magnitudes are listed in Table \ref{tab04ab_sec_std}. North is up and east to the left. The field of view is 10 $\times$ 10 arcmin$^2$.}
\label{fig04ab_field}
\end{figure}

In this paper, we present optical $BVRI$ photometric and medium-resolution spectroscopic analysis of a highly reddened type Ia supernova SN 2004ab. 
It was discovered on 2004 February 21.98 ({\sc ut}),
at 2 arcsec west and 11 arcsec north (refer Fig. \ref{fig04ab_field}) of the center of NGC 5054 \citep{mona04_CBET61}.  NGC 5054 is an SA(s)bc type galaxy at a  redshift of $z$ = 0.0058 \citep[][source NED]{pisa11}. SN 2004ab was classified as a type Ia supernova on February 24 and suggested to be a highly reddened supernova, caught about one week after maximum light. The expansion velocity measured using absorption minimum of  Si\,{\sc ii} $\lambda$6355 was  10400 km\,s$^{-1}$ \citep{math04}. 

The paper is organized as follows. Section \ref{sec04ab_observation} describes the observation and data reduction techniques. Photometric results are presented in Section \ref{sec04ab_results}. Anomalous extinction of host galaxy NGC 5054 is discussed in Section \ref{sec04ab_reddening}. Absolute and bolometric luminosities are estimated in Section \ref{sec04ab_absolute}. 
Spectroscopic results are presented in Section \ref{sec04ab_spec_results}.
The paper is summarized in Section \ref{sec04ab_summary}.

\section{OBSERVATIONS AND DATA REDUCTION}
\label{sec04ab_observation}

\subsection{Imaging}
Observations of SN 2004ab were carried out using the Himalaya Faint Object Spectrograph Camera (HFOSC) attached to the 2-m Himalayan Chandra Telescope (HCT) of Indian Astronomical Observatory (IAO), Hanle, India. HFOSC is equipped with a 2K$\times$4K pixels, SITe CCD chip. The central 2K$\times$2K pixels of the chip with a field of view of 10 $\times$ 10 arcmin$^2$,  at a plate scale of 0.296 arcsec\,pixel$^{-1}$ was used for imaging observations. The HFOSC CCD has gain of 1.22 electron\,ADU$^{-1}$ and readout noise of 4.87 electrons. 

The photometric monitoring of SN 2004ab began on 2004 February 24 and continued till 2004 June 22 in Bessell's $B$, $V$, $R$ and $I$ filters. 
 \citet{land92} standard star fields were observed on 2004 February 24 (PG0918+029, PG0942-029, SA101, SA104), March 01 (PG0918+029, PG1047+003, PG1323-086, PG1530+057), March 10 (PG0918+029, PG0942-029, PG1047+003, PG1323-086) and May 11 (PG1633+099, PG1657+078, SA107) under photometric conditions. These fields were used to estimate the atmospheric extinction and transformation co-efficients, and to calibrate a sequence of secondary standards in the supernova field.  Standard fields were monitored  in the airmass range $\sim$1.1--2.0 for estimting atmospheric extinction whereas for  determining   transformation co-efficients we restricted to  airmass range of $\sim$1.1--1.4.
   
The CCD images were processed using standard {\sc iraf}\footnote {{\sc iraf} is distributed 
by the National Optical Astronomy Observatories, which are operated by the 
Association of Universities for Research in Astronomy, Inc., under cooperative 
agreement with the National Science Foundation} routines. The images were bias corrected with a median combined master bias frame obtained using all the bias frames taken throughout the night. Flat-field correction was done using the median combined  normalized flat-field images of the twilight sky in different bands. Cosmic ray hits were removed from the flat-field corrected images.

 Aperture photometry was performed on the stars of Landolt standard fields 
using  DAOPHOT package of {\sc iraf}.
Bright stars in the Landolt standard field were used to determine the aperture 
growth curve and compute aperture corrections by measuring the magnitude difference 
at an aperture radius $\sim$ 3-4 times the full width half maximum (FWHM) and
 at an aperture close to the FWHM of the stellar profile. Magnitudes of Landolt 
standard stars  were obtained by applying the aperture corrections to the magnitude 
determined at a radius (close to the FWHM) that maximized the signal-to-noise (S/N) ratio.  
The nightly extinction co-efficients in different bands were determined.  The observed 
magnitudes of the Landolt standards were corrected for atmospheric  extinction using 
the estimated nightly extinction co-efficients. 
Finally, the corrected magnitudes of Landolt standards were used to derive 
the colour terms and photometric zero points on each night. 

 We selected a sequence of stars in the field of SN 2004ab (marked in Fig. 
\ref{fig04ab_field}) to use as secondary standards. Aperture photometry, 
as discussed above, was performed on these stars  on the nights when Landolt standard 
stars were obeserved. Magnitudes of this sequence of stars  were determined by 
applying aperture corrections computed using bright stars in the field of SN 2004ab. 
The observed magnitudes were  extinction corrected using the nightly extinction 
co-efficients.  The extinction corrected magnitudes were then calibrated  using  
colour terms and zero points obtained using observations of  Landolt standards.

The average $BVRI$ magnitudes of the sequence of secondary standards in the 
field of SN 2004ab are given in Table \ref{tab04ab_sec_std}.  The errors associated 
with  magnitudes of secondary standards are the standard deviation of the 
magnitudes obtained on the four calibration nights. These secondary standards 
were used to calibrate the supernova magnitude obtained on other nights.

SN 2004ab occurred close to the nucleus of the host galaxy,  in a region with  very high and highly varying background (refer Fig. \ref{fig04ab_field}). The contamination of galaxy background prohibits  proper estimation  of supernova magnitude. Hence, template subtraction method was used to subtract the galaxy background. The $BVRI$ template frames were prepared from the multiple deep exposures of host galaxy NGC 5054, obtained with the same instrumental setup and  under  good seeing condition on 2005 May 28.
Template frames were subtracted from the supernova frames of respective bands. This leaves behind only the  supernova.  Magnitudes of the supernova and secondary standards were  extracted by aperture photometry with a smaller aperture radius ($\sim$ FWHM). The aperture correction,  determined using bright stars in the supernova field was  applied to the extracted magnitudes of supernova and secondary standards.  The nightly  zero-points,   determined from the observed and calibrated magnitudes of secondary standards and average of derived colour terms for four nights,  were used to  calibrate the supernova magnitudes. As the reddening in the direction of SN 2004ab  is high (refer Section \ref{sec04ab_reddening}), effect of second order extinction was taken into account in all the calibration process. The derived supernova magnitudes are listed in Table \ref{tab04ab_sn_mag}. The final error in the supernova  magnitudes was estimated by adding in quadrature the photometric error at an aperture  with maximum S/N ratio (i.e.  photometric error computed
by {\sc iraf}), error in aperture correction, and the error associated with nightly photometric zero-points.

\begin{table*}
\centering
\caption[]{Magnitudes for the sequence of secondary standard stars in
the field of SN 2004ab. The stars are marked in Fig. \ref{fig04ab_field}}
\begin{tabular}{@{}lcccc@{}}
\hline
ID & B & V &  R & I\\
\hline 
1 &14.773 $\pm$	0.033&14.055 $\pm$ 0.017&13.637 $\pm$ 0.017&13.246 $\pm$ 0.005\\
2 &15.144 $\pm$ 0.035&14.227 $\pm$ 0.016&13.667 $\pm$ 0.015&13.111 $\pm$ 0.009\\
3 &14.875 $\pm$ 0.032&14.241 $\pm$ 0.013&13.851 $\pm$ 0.018&13.475 $\pm$ 0.010\\
4 &15.175 $\pm$ 0.031&14.347 $\pm$ 0.013&13.867 $\pm$ 0.016&13.421 $\pm$ 0.017\\
5 &15.343 $\pm$ 0.034&14.693 $\pm$ 0.020&14.290 $\pm$ 0.023&13.921 $\pm$ 0.024\\
6 &16.044 $\pm$ 0.024&15.226 $\pm$ 0.010&14.753 $\pm$ 0.012&14.306 $\pm$ 0.011\\
7 &15.954 $\pm$ 0.037&15.372 $\pm$ 0.017&15.001 $\pm$ 0.016&14.627 $\pm$ 0.004\\
8 &17.518 $\pm$ 0.048&16.082 $\pm$ 0.013&15.119 $\pm$ 0.017&14.139 $\pm$ 0.011\\
9 &16.916 $\pm$ 0.021&16.326 $\pm$ 0.017&15.953 $\pm$ 0.014&15.581 $\pm$ 0.024\\
10&17.106 $\pm$ 0.022&16.456 $\pm$ 0.015&16.047 $\pm$ 0.023&15.665 $\pm$ 0.008\\
11&17.640 $\pm$ 0.058&16.557 $\pm$ 0.014&15.873 $\pm$ 0.016&15.278 $\pm$ 0.003\\
\hline
\end{tabular}
\label{tab04ab_sec_std}
\end{table*}

\begin{figure}
\centering
\resizebox{\hsize}{!}{\includegraphics{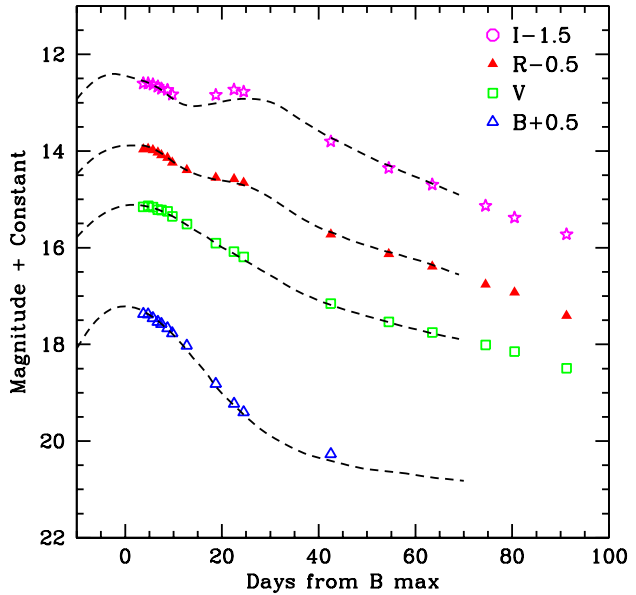}}
\caption[]{$BVRI$ light curves of SN 2004ab. The light curves have been shifted by the amount indicated in the legend. The phase is measured in days from the $B$ band maximum. The dashed lines represent template light curves obtained using SNooPy code.}
\label{fig04ab_lc}
\end{figure}
 
\begin{table*}
\centering
\caption{Optical $BVRI$ photometry of SN 2004ab with HCT.}
\begin{tabular}{@{}lcrcccc@{}}
\hline
 Date & JD$^a$ & Phase$^b$ & B & V & R & I\\
\hline
24/02/2004&3060.44&3.62 &16.867 $\pm$ 0.008  &15.154 $\pm$ 0.006  &14.461 $\pm$ 0.011  &14.100 $\pm$ 0.014\\        
25/02/2004&3061.49&4.67 &16.880 $\pm$ 0.011  &15.133 $\pm$ 0.007  &14.450 $\pm$ 0.012  &14.100 $\pm$ 0.016\\        
26/02/2004&3062.41&5.59 &16.958 $\pm$ 0.011  &15.161 $\pm$ 0.014  &14.479 $\pm$ 0.011  &14.122 $\pm$ 0.008\\        
27/02/2004&3063.38&6.56 &17.031 $\pm$ 0.009  &15.213 $\pm$ 0.009  &14.532 $\pm$ 0.006  &14.166 $\pm$ 0.017\\        
28/02/2004&3064.35&7.53 &17.078 $\pm$ 0.007  &15.219 $\pm$ 0.008  &14.586 $\pm$ 0.004  &14.210 $\pm$ 0.013\\        
29/02/2004&3065.40&8.58 &17.167 $\pm$ 0.008  &15.246 $\pm$ 0.005  &14.638 $\pm$ 0.005  &14.235 $\pm$ 0.011\\        
01/03/2004&3066.42&9.60 &17.269 $\pm$ 0.009  &15.349 $\pm$ 0.008  &14.739 $\pm$ 0.010  &14.330 $\pm$ 0.010\\        
04/03/2004&3069.39&12.57&17.528 $\pm$ 0.032  &15.512 $\pm$ 0.011  &14.890 $\pm$ 0.013  &                  \\        
10/03/2004&3075.39&18.57&18.315 $\pm$ 0.043  &15.906 $\pm$ 0.017  &15.048 $\pm$ 0.011  &14.340 $\pm$ 0.018\\        
14/03/2004&3079.28&22.46&18.726 $\pm$ 0.026  &16.079 $\pm$ 0.010  &15.079 $\pm$ 0.017  &14.229 $\pm$ 0.020\\        
16/03/2004&3081.33&24.51&18.903 $\pm$ 0.033  &16.189 $\pm$ 0.014  &15.155 $\pm$ 0.013  &14.274 $\pm$ 0.017\\        
03/04/2004&3099.26&42.44&19.772 $\pm$ 0.066  &17.157 $\pm$ 0.016  &16.226 $\pm$ 0.014  &15.302 $\pm$ 0.014\\        
15/04/2004&3111.26&54.44&                    &17.532 $\pm$ 0.009  &16.629 $\pm$ 0.009  &15.852 $\pm$ 0.010\\        
24/04/2004&3120.31&63.49&                    &17.752 $\pm$ 0.013  &16.890 $\pm$ 0.017  &16.195 $\pm$ 0.025\\        
05/05/2004&3131.22&74.40&                    &18.014 $\pm$ 0.017  &17.262 $\pm$ 0.017  &16.634 $\pm$ 0.025\\        
11/05/2004&3137.23&80.41&                    &18.147 $\pm$ 0.022  &17.427 $\pm$ 0.019  &16.879 $\pm$ 0.026\\        
22/05/2004&3148.12&91.30&                    &18.493 $\pm$ 0.021  &17.905 $\pm$ 0.028  &17.225 $\pm$ 0.022\\        
\hline
\multicolumn{7}{@{}l}{$^a$2450000+; $^b$Observed phase in days with respect to the epoch of $B$ band maximum: JD = 245 3056.82}
\end{tabular}
\label{tab04ab_sn_mag}
\end{table*}

\subsection{Spectroscopy}
\label{sec04ab_spectroscopy}  
Spectroscopic observations of SN 2004ab were carried out from 2004, February 24 to May 05 on  11 occassions. Spectra were obtained  using Gr\#7 (wavelength range 3500--7800 \AA) and Gr\#8 (wavelength range 5200--9250 \AA) of HFOSC at a spectral resolution of $\sim$7 \AA. Log of spectroscopic observations is given in Table \ref{tab04ab_spec_log}. Spectrophotometric standard stars Feige 34, Feige 66, BD +33$^\circ$ 2642, HZ 44 and Wolf 1346 were observed for flux calibration. All the spectral data were processed in a standard manner using various tasks of {\sc iraf}. One-dimensional spectra were extracted using the optimal extraction method. The dispersion solutions obtained using arc lamp spectra of FeNe and FeAr were used to wavelength calibrate the supernova spectra.
Wavelength calibrated spectra were cross checked using bright night sky emission lines and wherever required,  small shifts were applied. Spectrophotometric standard stars observed with broader slit on the same night or nearby nights were used to correct for the instrumental response and atmospheric extinction using average spectroscopic extinction curve for the site. The spectra obtained in Gr\#7 (blue region) and Gr\#8 (red region) were combined by scaling one of the spectra to  the weighted mean of the other in the overlapping region,  to get a single spectrum covering blue to red region. 
 We performed bandpass spectrophotometry of the spectra using the transmissions curves of filters available with the HFOSC. The photometric flux  obtained  using available photometry is compared with the spectroscopic flux within a given  bandpass, and the scale factor was determined in each band. A cubic spline curve going through the scale factor in each band was fit and the spectra were scaled using this curve to match photometric and spectroscopic fluxes.
This way the combined spectra were brought to an absolute flux scale.  The spectra were then dereddened and corrected  for the redshift of the host galaxy NGC 5054, using $z$ = 0.0058 \citep[][source NED]{pisa11}.

\section{Photometric results}
\label{sec04ab_results}

The $BVRI$ light curves of SN 2004ab are presented in Fig. \ref{fig04ab_lc}. SN 2004ab was discovered post-maximum and our monitoring started $\sim$3 d  after the discovery. To determine the date of maximum and  brightness at maximum, we used the SNooPy \citep{burn11} software package, developed by the Carnegie Supernova Project (CSP),  for analysing  light curves  of SNe Ia. 

The  best matching templates to the observed $BVRI$ light curves of SN 2004ab,   generated using SNooPy  are  shown in Fig. \ref{fig04ab_lc} alongwith the observed light curves. The date of maximum and peak brightness in different  bands, obtained using  template fit are listed in Table \ref{tab04ab_lc_parameter}.  
The maximum in $V$, $R$ and $I$ band occurred $\sim$+1.5, +2.8 and $-$2 d,  relative to $B$ band maximum. This is in accordance with the observed
trend in  normal SNe Ia, where maximum in $I$ band precedes and  in $V$ and $R$ band  follows  maximum in $B$ band. The maximum brightness in $B$, $V$, $R$ and $I$ band is estimated as 16.71, 15.11, 14.38 and 13.90 magnitudes, respectively. Similar to normal SNe Ia, SN 2004ab also shows a shoulder in $R$ band and a pronounced secondary maximum in $I$ band. The secondary maximum  in  $I$ band is  $\sim$0.5 mag fainter with respect to first maximum and  occurred after $\sim$25 d.

\begin{table}
\centering
\caption{Log of spectroscopic observations of SN 2004ab.}
\begin{tabular}{@{}lcrc@{}}
\hline
Date & JD$^a$ & Phase$^b$& Range (\AA) \\
\hline
24/02/2004&3060.50&3.68  &3500--7800\\
25/02/2004&3061.50&4.68  &3500--7800\\
26/02/2004&3062.43&5.61  &3500--7800\\
28/02/2004&3064.37&7.55  &3500--7800; 5200--9250\\
29/02/2004&3065.35&8.53  &3500--7800; 5200--9250\\
04/03/2004&3069.36&12.54 &3500--7800; 5200--9250\\
08/03/2004&3073.46&16.64  &3500--7800; 5200--9250\\
14/03/2004&3079.33&22.51  &3500--7800; 5200--9250\\
03/04/2004&3099.33&42.51  &3500--7800; 5200--9250\\
26/04/2004&3122.20&65.38  &3500--7800; 5200--9250\\
05/05/2004&3131.25&74.43  &3500--7800; 5200--9250\\
\hline        
\multicolumn{4}{@{}l@{}}{$^a$2450000+; $^b$in days relative to $B$ band maximum.}
\end{tabular}
\label{tab04ab_spec_log}
\end{table}

The decline rate parameter, $\Delta m_{15}(B)$ for SN 2004ab is estimated as 1.17. Decline rate parameters  and rate of light curve decline during the late phase (40--90 d) in $VRI$ bands estimated by least square fit to the observed data are listed in Table \ref{tab04ab_lc_parameter}.  \citet{phil99} have shown that reddening acts to decrease the decline rate.  Hence, the measured  decline rate parameter needs to  be corrected for reddening. The observed decline rate parameter translates to the intrinsic decline rate parameter  $\Delta m_{15}(B)_\text{true}$ = 1.27,  after correcting for reddening (refer Section \ref{sec04ab_reddening}) using the updated relation by \citet{fola10}. 

\begin{figure*}
\centering    
\resizebox{0.7\hsize}{!}{\includegraphics{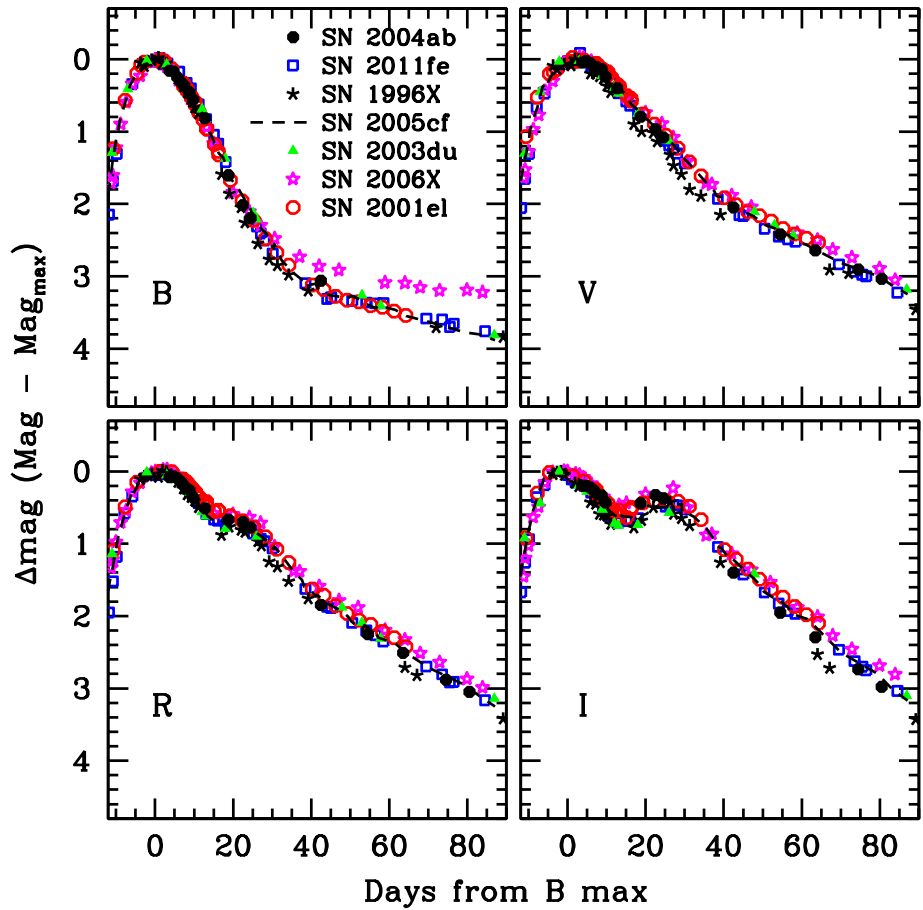}}
\caption[]{$BVRI$ light curves of SN 2004ab compared with those of SN 1996X, SN 2001el,  SN 2003du, SN 2005cf, SN 2006X and SN 2011fe. All the light curves have been shifted to match their peak magnitudes and to the epoch of $B$ band maximum.}
\label{fig04ab_lc_comp}
\end{figure*}

In Fig. \ref{fig04ab_lc_comp}, the $BVRI$ light curves of SN 2004ab are compared with those of 
SN 1996X ($\Delta m_{15}(B)$ =1.31; \citealt{salv01}), 
SN 2001el ($\Delta m_{15}(B)$ = 1.13; \citealt{kris03}), 
SN 2003du ($\Delta m_{15}(B)$ = 1.04; \citealt*{anup05,stan07}),
SN 2005cf ($\Delta m_{15}(B)$ = 1.12; \citealt{past07,wang09}) and
SN 2006X ($\Delta m_{15}(B)$ =1.31; \citealt{wang08}),
SN 2011fe ($\Delta m_{15}(B)$ = 1.07; \citealt{vink12,rich12}).
All the light curves have been shifted to match their peak magnitude in the respective bands and to the epoch of $B$ band maximum. The $B$ band light curve of SN 2004ab  closely resembles those of SN 2001el, SN 2003du, SN 2005cf and SN 2011fe (till $\sim$25 d). In $V$ and $R$ band, the light curve of SN 2004ab is similar to those of SN 2003du, SN 2005cf and SN 2011fe.   The early phase light curve of SN 2004ab in $I$ band looks similar to that of SN 2001el. After  secondary maximum, SN 2004ab is  fainter than SN 2001el. Light curve of SN 2004ab is wider than SN 1996X in  all the bands. 

\begin{table*}
\centering
\caption{Photometric parameters of SN 2004ab.}
\small
\begin{tabular}{@{}lcccccc@{}}
\hline
Band & JD (max)$^a$  & $m_{\lambda}^\text{max}$& $A_\lambda$ &$M_{\lambda}^{\text{max}^b}$&  $\Delta m_{15}(\lambda)$ &Decline Rate$^c$ \\  
    &                &                         & Total &             &             & during 40--90 d \\
\hline
$B$ &3056.8 $\pm$ 0.5&16.71 $\pm$ 0.05&4.15 $\pm$ 0.19&$-$19.31 $\pm$ 0.25 &1.17 $\pm$ 0.05&  -  \\
$V$ &3058.3 $\pm$ 0.5&15.11 $\pm$ 0.04&2.51 $\pm$ 0.11&$-$19.28 $\pm$ 0.19 &0.64 $\pm$ 0.04&2.645\\
$R$ &3059.6 $\pm$ 0.5&14.38 $\pm$ 0.05&1.82 $\pm$ 0.08&$-$19.31 $\pm$ 0.18 &0.70 $\pm$ 0.05&3.352\\
$I$ &3054.8 $\pm$ 0.5&13.90 $\pm$ 0.05&1.00 $\pm$ 0.04&$-$18.97 $\pm$ 0.16 &0.66 $\pm$ 0.05&3.950\\
\hline
\multicolumn{7}{@{}l}{$^a$245 0000+; $^b$For $\mu$ =  31.87 and $R_V\text{(host)}$ = 1.41}\\
\multicolumn{7}{@{}l}{$^c$In unit of mag\,(100 d)$^{-1}$ and epoch is relative to $B$ band maximum.} 
\end{tabular}
\label{tab04ab_lc_parameter}
\end{table*}

\section{Anomalous extinction of host galaxy towards SN 2004\MakeLowercase{ab}}
\label{sec04ab_reddening}

SN 2004ab occurred very close to the nucleus of the host galaxy NGC 5054. Hence, a substantial amount of reddening due to ISM within the host galaxy is expected. The spectra of SN 2004ab show strong narrow Na\,{\sc i} D absorption lines (refer Section \ref{sec04ab_spec_results}) at the rest frame of host galaxy with an average equivalent width (EW) of 3.3 \AA. \citet{math04} also reported similar EW of narrow Na\,{\sc i} D lines. The measured EW gives  reddening of $E(B-V)_\text{host}$ = 0.53 and 1.68 mag, on using the two relations of \citet*{tura03} between  EW of Na\,{\sc i} D lines and $E(B-V)$. 

Reddening suffered by SN 2004ab is also estimated using various photometric methods. The relations  between the observed SN colour at maximum and $\Delta m_{15}(B)$ \citep{phil99, alta04} give total reddening of 1.65 mag. The relation of \citet{wang06} between the $(B-V)$ colour measured at 12 d after the $B$ band maximum (referred as $\Delta C_{12}$) and decline rate parameter gives $E(B-V)_\text{total}$ as 1.66 mag.
The method of \citet{rein05} gives total reddening of 1.60 mag at maximum and 1.70 mag at 35 d after maximum. 
Average of all the reddening values derived using photometric methods is 1.65 $\pm$ 0.04 mag. 

\begin{figure*}
\centering    
\resizebox{0.7\hsize}{!}{\includegraphics{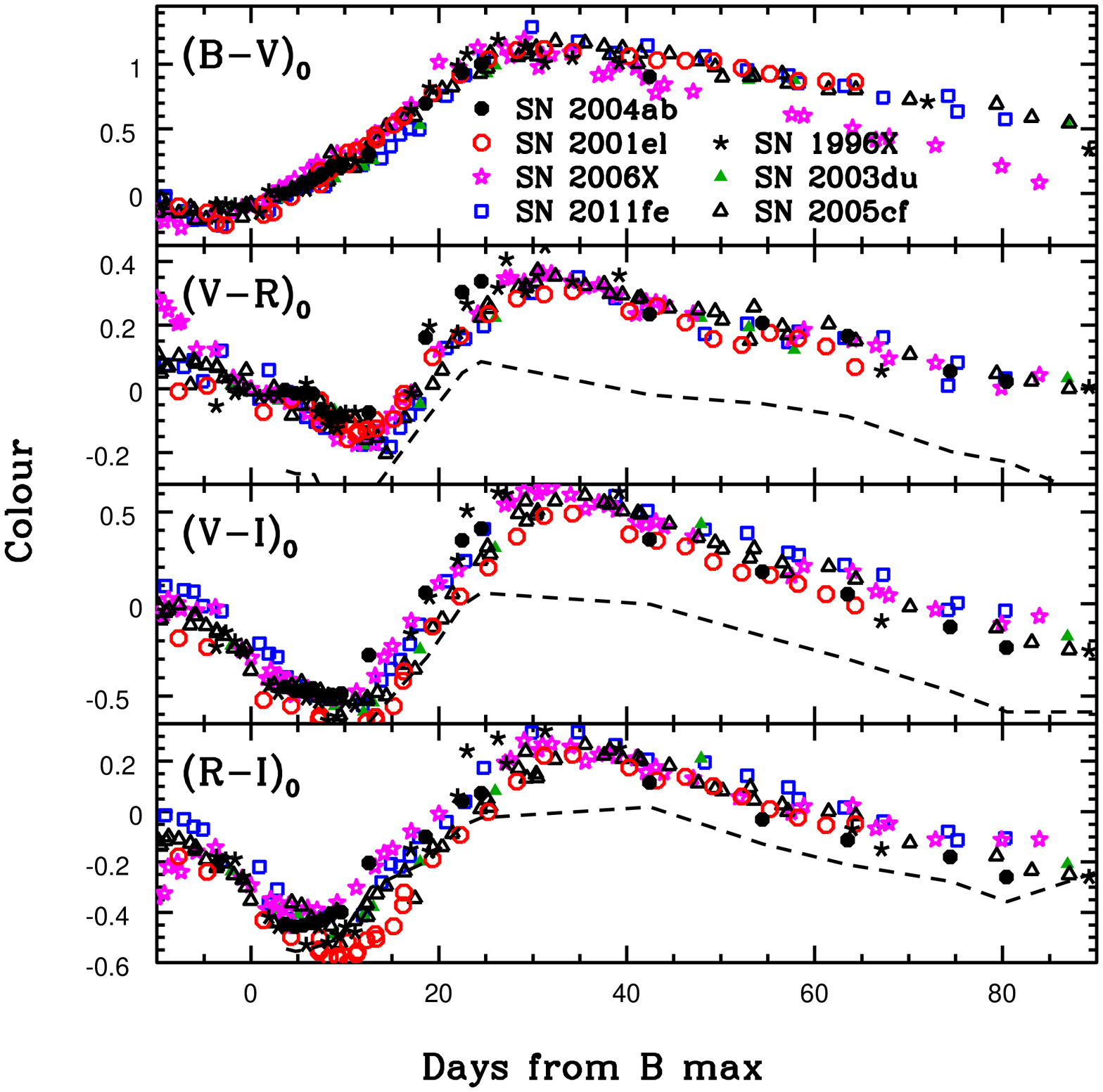}}
\caption[]{The dereddened $(B-V)$, $(V-R)$, $(V-I)$ and $(R-I)$ colour curves of SN 2004ab plotted with those of some well studied SNe Ia. The colour curves of SN 2004ab are corrected for $E(B-V)_\text{Gal}$ = 0.07 mag, $E(B-V)_\text{host}$ = 1.63 mag and $R_V\text{(host)}$ = 1.39 using CCM model with $a_\lambda$ and $b_\lambda$ coefficients from \citet{kris06}. The dashed lines represent dereddened colour curves of SN 2004ab if coefficients are used from \citet{card89}. }
\label{fig04ab_colour}
\end{figure*}

Contribution of reddening due to ISM within our Galaxy is $E(B-V)_\text{Gal}$ = 0.07 \citep*{schl98,schl11}. As an alternative method to estimate  reddening suffered by SN 2004ab, we compared the observed $(B-V)$ colour of SN 2004ab with reddening corrected $(B-V)$ colour curves of well studied SNe Ia (refer Fig. \ref{fig04ab_colour}). The colour curves are dereddened using the reddening values mentioned in the respective references. The $(B-V)$ colour curve of SN 2004ab was then shifted to match with the colour curves of other SNe used in comparison. A $\chi^2$ minimization was used to estimate the offset between  $(B-V)$ colour curve of SN 2004ab and those of other SNe.  
After   shifting the   $(B-V)$ colour curve of SN 2004ab by 1.70 $\pm$ 0.05 mag, it matches well with those of other SNe (refer Fig. \ref{fig04ab_colour}). This value of colour excess is consistent with the $E(B-V)_\text{total}$ derived using photometric and spectroscopic methods. For further analysis we have used $E(B-V)_\text{total}$ = 1.70 mag as total reddening for SN 2004ab. 

From the colour excess $E(B-V)$, the extinction in $V$ band $A_V$ is estimated using the following relation,
\begin{equation}
 A_V = R_V \times E(B-V)
\end{equation}
where $R_V$ is the ratio of total-to-selective extinction. Using the standard Galactic value of $R_V$ = 3.1 and the estimated value of $E(B-V)$ = 1.70 mag, the above relation gives total extinction in $V$ band as $A_V$ = 5.27 mag.  

NGC 5054, the host galaxy of SN 2004ab,  has a radial velocity corrected for local group 
infall onto Virgo cluster of 1704 km\,s$^{-1}$ (LEDA database). Assuming 
$H_{0}$ = 72 km\,s$^{-1}$\,Mpc$^{-1}$ \citep{free01}, we derived distance modulus of $\mu$ = 31.87 $\pm$ 0.15 mag for  NGC 5054. This leads to  $V$ band peak absolute magnitude of SN 2004ab,     $M_V^\text{max}$ = $-$22.03 $\pm$ 0.15 mag, making it brighter by more than 2 magnitudes than the normal SNe Ia. 
This indicates that  $R_V$ = 3.1 is not applicable to the host galaxy component of extinction for SN 2004ab.

\subsection{Estimation of $R_V$ using photometry}
 Several studies  have suggested  lower values of $R_V$ than its canonical value of 3.1 towards SNe Ia in their host galaxies \citep{elia06,kris06,kris07,wang08,fola10,aman14}. This indicates that the extinction properties of dust in the host galaxies, towards SNe Ia is different from those of Milky Way. The lower value of $R_V$ is an indicator of  smaller  dust grains compared to  that of the  Milky Way. This type of dust is generally referred as non-standard dust and the reddening resulting from it as  non-standard reddening. 
The  extinction due to dust in our Galaxy  is  explained with the  model of \citet*{card89} commonly known as CCM model. Extinction in this model is parametrized by the following relation, 

\begin{equation}
 \frac{A_\lambda}{A_V} = a_\lambda + \frac{b_\lambda}{R_V}
\end{equation}

where $A_V = R_V \times E(B-V)$, $a_\lambda$ and $b_\lambda$ are wavelength-dependent co-efficients. 
\citet{kris06} have suggested that the  co-efficients of reddening model for host galaxies of SNe Ia,  should be derived from  SED of SNe rather than from the SED of normal stars. They have derived  values of  $a_\lambda$ and $b_\lambda$ for $UBVRI$ bands using  type Ia spectral template of \citet*{nuge02}.
To estimate $R_V$  towards  SN 2004ab in the host galaxy, we used dereddened colour curves of well studied SNe Ia. The reddening corrected $(B-V)$, $(V-R)$, $(V-I)$ and $(R-I)$ colour curves of these SNe are plotted in Fig. \ref{fig04ab_colour}. 

The colour curves of SN 2004ab are first corrected for the Galactic extinction of $E(B-V)_\text{Gal}$ = 0.07 mag, and  then  for $E(B-V)_\text{host}$ = 1.63 using CCM extinction law with varying $R_V$.   A $\chi^2$ minimization was used to estimate the value of $R_V$ using which  the dereddened $(V-R)$, $(V-I)$ and $(R-I)$ colour curves of SN 2004ab match with  those of  well studied SNe Ia. It is found that with $R_V$ = 1.39 $\pm$ 0.05 and values of $a_\lambda$,  $b_\lambda$ as derived by  \citet{kris06}, the dereddened colour curves of SN 2004ab match well with colour curves of normal SNe Ia used in comparison.  Using this analysis we derived $E(V-R)$ = 0.64 $\pm$ 0.06 mag, $E(V-I)$ = 1.39 $\pm$ 0.09 mag and $E(R-I)$ = 0.75 $\pm$ 0.07 mag. The values of $a_\lambda$ and $b_\lambda$  from \citet{card89} give poor fit (shown by dashed lines in Fig. \ref{fig04ab_colour}). The derived value of $R_V$ = 1.39 towards SN 2004ab in  the host galaxy is smaller than the canonical value of $R_V$ = 3.1 for Milky Way. There are many host galaxies showing smaller value of $R_V$ towards SNe Ia,  some of them are listed in Table \ref{tab04ab_Rv}. 

\begin{figure}
\centering    
\resizebox{\hsize}{!}{\includegraphics{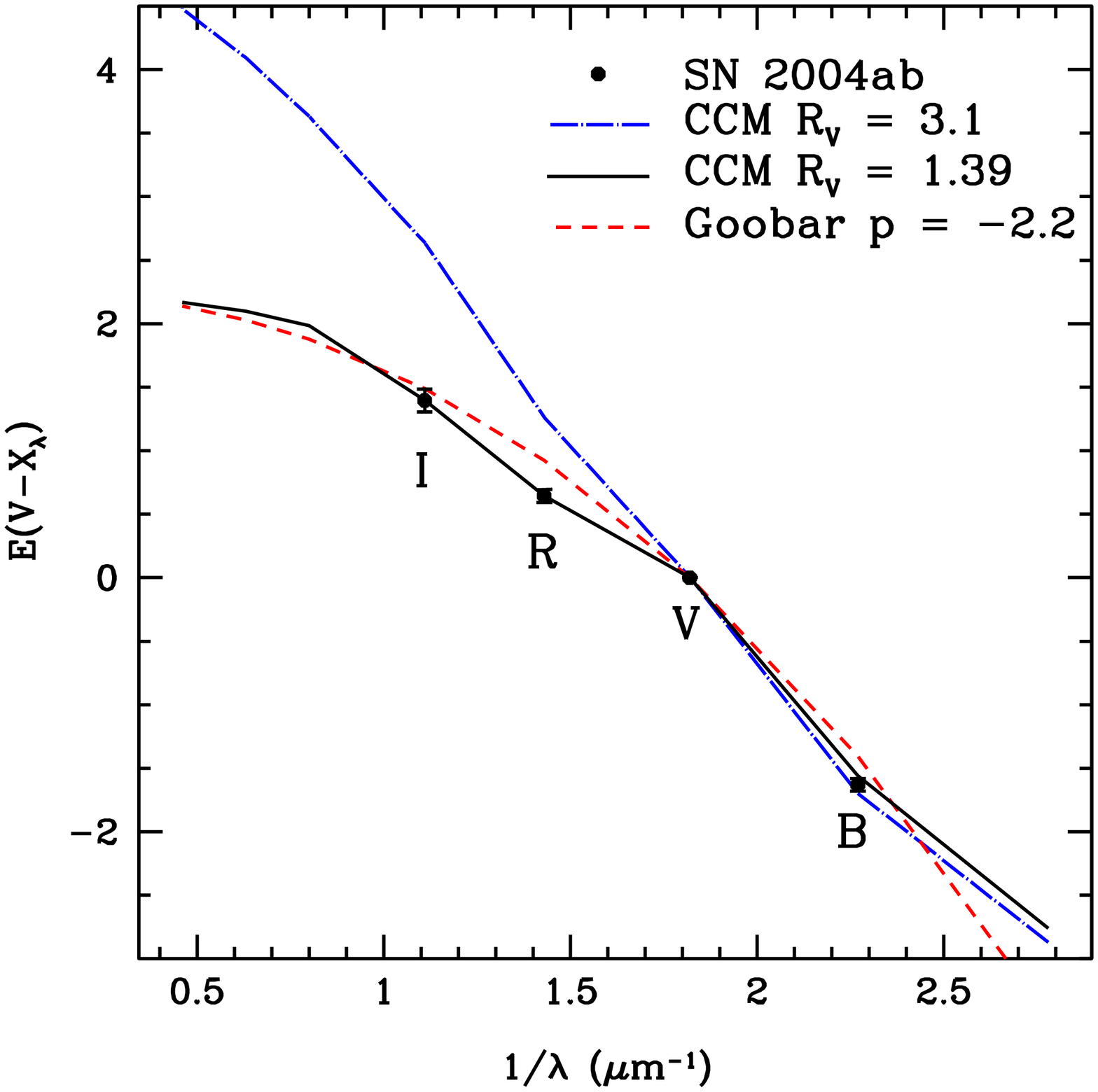}}
\caption[]{Colour excesses $E(V-X_\lambda)$, where $X_\lambda$ = $BVRI$ measurement for SN 2004ab. CCM extinction model $A_\lambda/A_V = a_\lambda + (b_\lambda/R_V)$
with $R_V$ = 3.1, $R_V$ = 1.39 and power law extinction model $A_\lambda/A_V$ = ($\lambda/\lambda_V)^p$ of \citet{goob08} with $p = -2.2$ are also displayed.}
\label{fig04ab_Rv}
\end{figure}

Alternative to CCM, the extinction model with power-law of type $A_\lambda/A_V$ = ($\lambda/\lambda_V)^p$,  expected from the  multiple scattering of light due to a dusty CSM is proposed \citep{goob08,aman14}. The multiple scattering of photons by circumstellar dust steepens the effective extinction law. The measured colour excess of SN 2004ab is fit with this power law. A  reasonably good fit is achieved for $p = -2.2$. The fit is shown in Fig. \ref{fig04ab_Rv} along with the CCM model with $R_V$ = 1.39 and $R_V$ = 3.1.

\subsection{Spectroscopic view of extinction}

\begin{figure}
\centering    
\resizebox{\hsize}{!}{\includegraphics{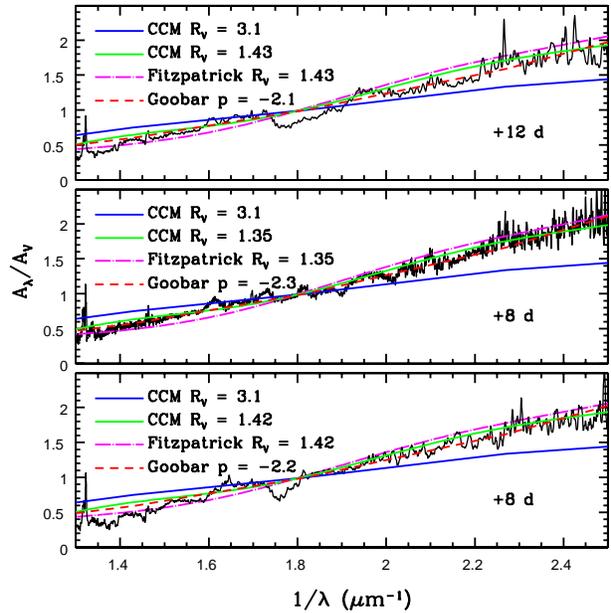}}
\caption[]{Extinction  curve for SN 2004ab obtained using the pair of spectra of SN 1994D--2004ab (top and bottom panel) and SN 1996X--2004ab (middle panel). The best fits using CCM extinction model $A_\lambda/A_V = a_\lambda + (b_\lambda/R_V)$, power law extinction model $A_\lambda/A_V$ = ($\lambda/\lambda_V)^p$ \citep{goob08} and \citet{fitz99} model are displayed. The standard Milky Way extinction law with $R_V$ = 3.1 clearly deviates for SN 2004ab.}
\label{fig04ab_Rv_spec}
\end{figure}

The anomalous extinction of SN 2004ab is also verified using spectroscopic method described by  \citet{elia06}.
In this method  optical Spectral Energy Distribution (SED) of a reddened SN is compared with those of unreddened reference SNe Ia, at similar epochs. For this, SN 1994D \citep{pata96} and SN 1996X \citep{salv01}, having similar decline rate parameters are selected as reference objects. All the spectra are dereddened for Galactic extinction and redshift corrected. After this, the reference spectra were scaled to the distance of SN 2004ab. Extinction as a function of wavelength $A_\lambda$ is derived using the formula

\begin{equation}
 A_\lambda = -2.5 \log\frac{f_\text{04ab}}{f^\text{scaled}_\text{ref}}
\end{equation}

where $f_\text{04ab}$ and $f^\text{scaled}_\text{ref}$ are the observed fluxes of SN 2004ab and the reference SN scaled to the distance of SN 2004ab,  respectively. The extinction  curve is obtained by normalizing the  derived extinction
$A_\lambda$ at the $V$ band effective wavelength. 
We used three pairs of SN 1994D--2004ab and two pairs of SN 1996X--2004ab, a total five  pairs of spectra for the analysis.  To derive the value of $R_V$, we fit CCM model to extinction curves of SN 2004ab. Three examples of fit are shown in Fig. \ref{fig04ab_Rv_spec}.  It is clear that CCM model with $R_V$ = 3.1  deviates for SN 2004ab,  whereas the model extinction curve with lower value of  $R_V$ ($\sim$ 1.4) fits the derived extinction curve reasonably well. The best fit values of $R_V$ are also listed in Fig. \ref{fig04ab_Rv_spec}.  The derived extinction curve is also consistent with the Fitzpatrick parametrization \citep{fitz99} of extinction curve with   $R_V$ $\sim$ 1.4.

 An attempt is made to fit the derived extinction curve of SN 2004ab with power law extinction model, $A_\lambda/A_V$ = ($\lambda/\lambda_V)^p$ of \citet{goob08}. The best fit model extinction curves and corresponding values of the power law index  $p$ are shown  in Fig. \ref{fig04ab_Rv_spec}. 
From the analysis of five pairs of spectra we estimate mean value of 
$R_V$ = 1.41 $\pm$ 0.03, which is similar to that derived from photometric method. 
Combining the results  obtained from spectroscopic (five pairs of spectra) and photometric methods, we derive $R_V$ = 1.41 $\pm$ 0.06. This value of $R_V$  is used in further analysis.

\begin{table*}
\caption[]{The host galaxies with low value of $R_V$ towards SNe Ia.}
\centering
\begin{tabular}{@{}lccccc@{}}
\hline
SN     &Host Galaxy & $\Delta m_{15}(B)_\text{true}$ & $E(B-V)_\text{host}$  & $R_V$ & Reference \\
\hline
1999cl & NGC 4501&1.29 $\pm$ 0.08 &1.24 $\pm$ 0.07 &1.55 $\pm$ 0.08 & \citet{kris06}    \\
2001el & NGC 1448&1.15 $\pm$ 0.04 &0.21 $\pm$ 0.05 &2.15 $\pm$ 0.23 & \citet{kris03}    \\
2003cg & NGC 3169&1.25 $\pm$ 0.05 &1.33 $\pm$ 0.11 &1.80 $\pm$ 0.19 & \citet{elia06}   \\
2005A  & NGC 958&1.34 $\pm$ 0.03 &1.11 $\pm$ 0.07 &1.68 $\pm$ 0.10 & \citet{fola10}              \\
2006X  & NGC 4321&1.31 $\pm$ 0.05 &1.42 $\pm$ 0.04 &1.48 $\pm$ 0.06 & \citet{wang08}         \\
2014J  & M82     &  1.08 $\pm$ 0.03 &  1.37 $\pm$ 0.03      & 1.40 $\pm$ 0.10 & \citet{aman14} \\
       &         &                  &                       &                 & \citet{shub16} \\
2004ab & NGC 5054&1.27 $\pm$ 0.05 &1.63 $\pm$ 0.05 &1.41 $\pm$ 0.06 & This work       \\
\hline
\label{tab04ab_Rv}
\end{tabular}
\end{table*}

\section{Absolute and  bolometric luminosity}
\label{sec04ab_absolute}

The  total extinction $A_\lambda$ suffered by SN 2004ab in $BVRI$ bands are derived (using $R_V^{host}$ = 1.41) as 4.15 $\pm$ 0.19, 2.51$\pm$ 0.11, 1.82 $\pm$ 0.08 and 1.00 $\pm$ 0.04 mag, respectively. Using these values and distance modulus of $\mu$ = 31.87 mag, the peak absolute magnitudes in different bands are calculated  and listed in Table \ref{tab04ab_lc_parameter}. 
  
Alternatively,  peak absolute magnitudes can also be estimated using  empirical {\it Luminosity decline rate relation}. Peak absolute magnitudes of type Ia SNe are known to correlate with $\Delta m_{15}(B)$ \citep{phil99}.  Using the calibration by \citet{fola10}, we derived $B$ band peak absolute magnitude of SN 2004ab as $-$19.24 $\pm$ 0.20 mag,  consistent with the value given in Table \ref{tab04ab_lc_parameter}.   

\begin{figure}
\centering
\resizebox{\hsize}{!}{\includegraphics{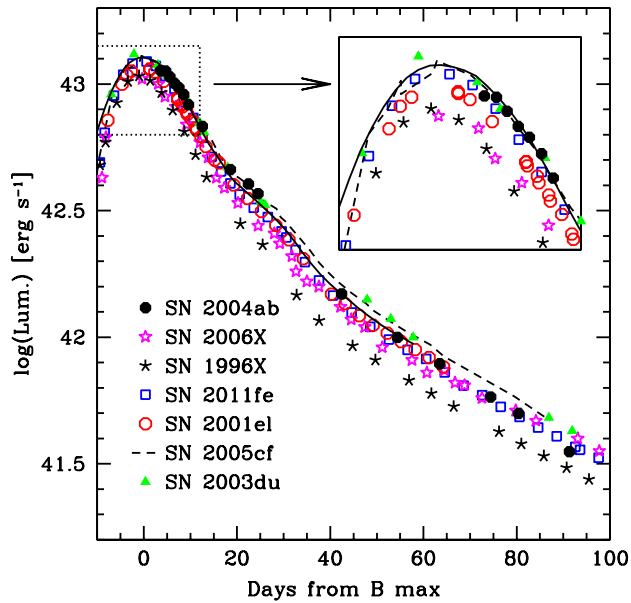}}
\caption[]{Quasi-bolometric light curve of SN 2004ab is plotted along with other well studied SNe Ia. The solid line represents quasi-bolometric light curve derived using the SNooPy fit to the observed data points. A zoomed view of plot around peak is shown in the inset}.
\label{fig04ab_bolo}
\end{figure}

The bolometric flux of SN 2004ab was obtained  using the observed $BVRI$ magnitudes listed in Table \ref{tab04ab_sn_mag}. The magnitudes were corrected  for total extinction. The extinction corrected  magnitudes were converted to flux using the zero points from \citet*{bess98}. A distance modulus of $\mu$ = 31.87 $\pm$ 0.15 mag was used to estimate the bolometric luminosity. Since SN 2004ab was observed in $BVRI$ bands, to account for the missing fluxes, we applied correction as described by \citet{wang09}. The quasi-bolometric light curve of SN 2004ab is plotted in Fig. \ref{fig04ab_bolo} and compared with those of other SNe Ia. The bolometric light curve of SN 2004ab is similar to those of SN 2001el and SN 2011fe. The peak bolometric luminosity of SN 2004ab is estimated  as $\log L_\text{bol}^\text{max}$ = 43.10 $\pm$ 0.07 erg\,s$^{-1}$. 

\subsection{Mass of Nickel synthesized}
The mass of $^{56}$Ni synthesized in the explosion of SN 2004ab is estimated using the Arnett's
rule \citep{arne82}. SN 2004ab was discovered on 2004 February 21.98. It was not detected  on 2004 February 1.15 upto a limiting mag of 18 mag  \citep{mona04_CBET61}. The bolometric light curve peaked on 2004 February 21.32. This indicates that  the rise time  $t_R <$ 20 d for SN 2004ab. The rise time of SNe Ia is found to correlate with $\Delta m_{15}(B)$. Brighter SNe with smaller $\Delta m_{15}(B)$ have longer rise time and fainter SNe with larger $\Delta m_{15}(B)$ have smaller rise time. 
The spectroscopically normal SNe Ia have typical rise time of 18--19.5 d \citep*{ries99,conl06,gane11}. Post-maximum decline rate and rise time relation of \citet{psko84} gives $t_R$ = 19 d for SN 2004ab. Using $t_R$ = 19 d, peak bolometric luminosity of $\log L_\text{bol}^\text{max}$ = 43.10 erg\,s$^{-1}$ and $\alpha$ = 1.2 \citep{bran92}, the mass of $^{56}$Ni synthesized in the explosion of SN 2004ab is estimated as M$_\text{Ni}$ = 0.53 $\pm$ 0.08 M$_\odot$.

\section{Spectroscopic results}
\label{sec04ab_spec_results}
\subsection{Spectral evolution}
\begin{figure}
\centering
\resizebox{\hsize}{!}{\includegraphics{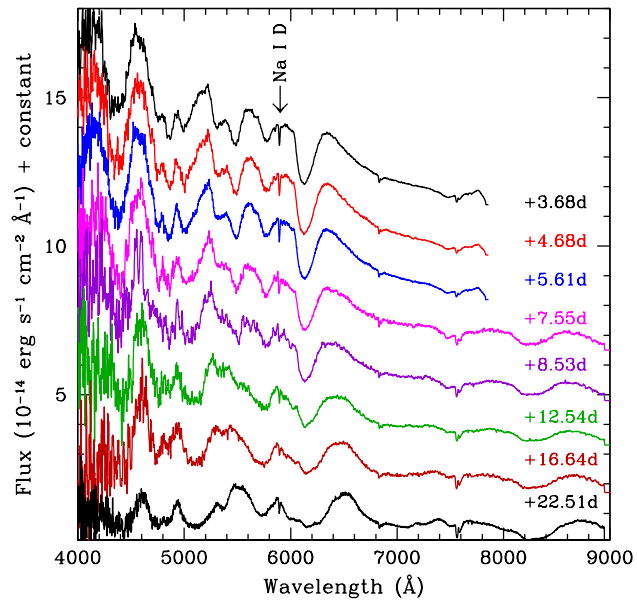}}
\caption[]{Spectral evolution of SN 2004ab from +3.7 to +22.5 d. Strong narrow Na\,{\sc i} D feature from host galaxy is clearly seen.}
\label{fig04ab_sp_p3p12}
\end{figure}

\begin{figure}
\centering
\resizebox{\hsize}{!}{\includegraphics{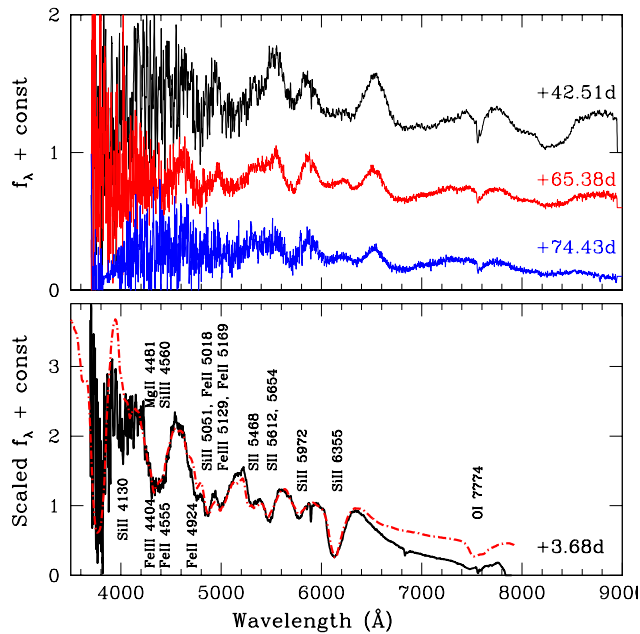}}
\caption[]{{\it Top:} Spectral evolution of SN 2004ab from +42.5 to +74.4 d. {\it Bottom:} The synthetic spectrum generated using {\sc syn++} code is compared with that of SN 2004ab at +3.7 d.}
\label{fig04ab_sp_p42p74}
\end{figure}

We obtained 11 spectra of SN 2004ab spanning from +3.7 to +74.4 d with respect to $B$ band maximum. The details of observations are given in Table \ref{tab04ab_spec_log}. Spectral evolution of SN 2004ab from +3.7 to +22.5 d is presented in Fig. \ref{fig04ab_sp_p3p12} and from +42.5 to +74.4 d in Fig. \ref{fig04ab_sp_p42p74}. All the spectra have been corrected for reddening (using $R_V^{host}$ = 1.41) as discussed in Section \ref{sec04ab_reddening} and redshift of $z$ = 0.0058.

The first spectrum of SN 2004ab obtained on  +3.7 d, shows features seen in normal SNe Ia. The  Fe\,{\sc iii} $\lambda$4404, Mg\,{\sc ii} $\lambda$4481, Fe\,{\sc ii} $\lambda$4555, and Si\,{\sc iii} $\lambda$4560 lines are blended giving a broad and deep profile in the region 4200--4600 \AA. During subsequent evolution, lines due to  Fe\,{\sc iii}, Si\,{\sc iii} weaken and the profile narrows down. Features of Fe\,{\sc ii} $\lambda$$\lambda$4924, 5018, Si\,{\sc ii} $\lambda$ 5051, Fe\,{\sc iii} $\lambda$5129 and Fe\,{\sc ii} $\lambda$5169 are clearly visible in the  4600--5200 \AA\ spectral region. The `W'-shaped S\,{\sc ii} $\lambda$$\lambda$5654, 5468, Si\,{\sc ii} $\lambda$5972 and Si\,{\sc ii} $\lambda$6355 characteristics of normal SNe Ia are strong in the spectrum of SN 2004ab. The strong narrow feature seen between Si\,{\sc ii} $\lambda$5972 and Si\,{\sc ii} $\lambda$6355 is due to Na\,{\sc i} D from host galaxy, indicating the high reddening within SN host. The overall appearance of  the first three spectra corresponding to  +3.7, +4.7 and +5.6 d is identical. The  spectrum obtained on  +7.6 d also covers the red region, where O\,{\sc i} and Ca\,{\sc ii} NIR triplet are clearly seen. The `W'-shaped S\,{\sc ii} lines are getting weaker. 
By +12.5 d, most of the Fe\,{\sc ii} lines have become stronger and S\,{\sc ii} feature is hardly visible. 
In the spectrum of +22.5 d, the Fe\,{\sc ii} lines are well developed,  the  Si\,{\sc ii} $\lambda$6355 is  getting replaced by a broad absorption profile due to increased contamination of Fe\,{\sc ii} lines. 

Spectrum at +42.5 d, shown in Fig. \ref{fig04ab_sp_p42p74} is dominated by strong  Fe\,{\sc ii} $\lambda$4924, $\lambda$5018, $\lambda$5169, $\lambda$5536, Na\,{\sc i}, and Ca\,{\sc ii} NIR triplet. Spectral appearance of the next two spectra obtained at +65.4 and +74.4 d is similar except for the weakening of Ca\,{\sc ii} NIR triplet. 

The spectrum of SN 2004ab at +3.7 d is fit with the synthetic spectrum generated using the 
{\sc syn++} code \citep*{thom11} and shown in Fig. \ref{fig04ab_sp_p42p74}.
The best fit to the observed spectrum 
is achieved at  photospheric velocity of 10700 km\,s$^{-1}$ and blackbody temperature of 14000 K. To reproduce the observed features, ions of O\,{\sc i}, Na\,{\sc i}, Mg\,{\sc ii}, Si\,{\sc ii}, Si\,{\sc iii}, S\,{\sc ii}, Ca\,{\sc ii}, Fe\,{\sc ii} and Fe\,{\sc iii}, each at an excitation temperature of 7000 K were included in the synthetic spectrum. Optical depths of each absorption feature was set to decrease exponentially with velocity, keeping {\it e}-folding velocity at 1000 km\,s$^{-1}$. The identified features are marked in  the spectrum. 

\begin{figure}
\centering
\resizebox{\hsize}{!}{\includegraphics[trim = 0mm 0mm 0mm 80mm, clip]{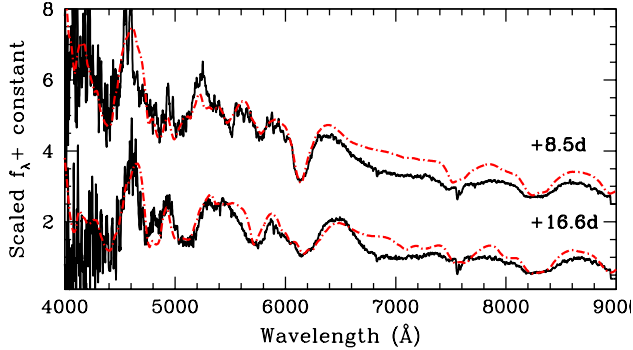}}
\caption[]{The synthetic spectra generated using {\sc syn++} code are compared with those of SN 2004ab at +8.5 and +16.6 d.}
\label{fig04ab_synow}
\end{figure}

The synthetic spectra  at +8.5 and +16.6 d are also generated and displayed in Fig. \ref{fig04ab_synow} along with the observed spectra. The best fit for spectrum at +8.5 d is obtained using photospheric velocity of 10500 km\,s$^{-1}$ and blackbody temperature of 13000 K. The ions included to reproduce the synthetic spectrum are same as that of +3.7 d. For spectrum at +16.6 d, photospheric velocity is set to 9500 km\,s$^{-1}$ and blackbody temperature at 10000 K. The synthetic spectrum includes ions of O\,{\sc i}, Na\,{\sc i}, Mg\,{\sc ii}, Si\,{\sc ii}, Ca\,{\sc ii} and Fe\,{\sc ii}.

\begin{figure}
\centering
\resizebox{\hsize}{!}{\includegraphics{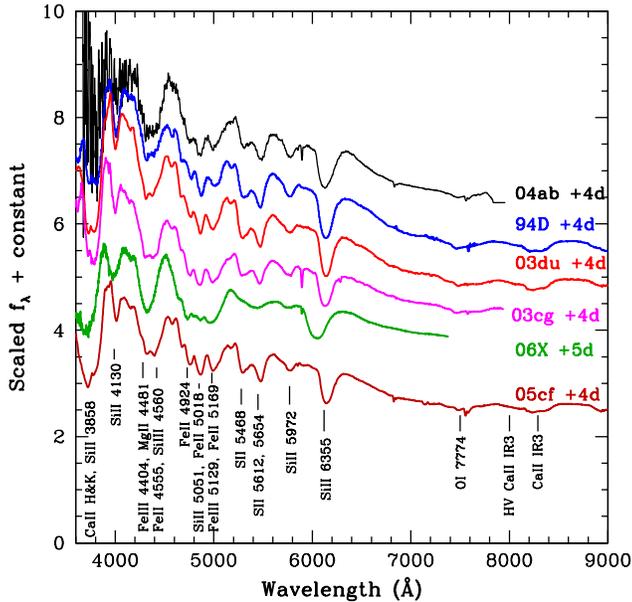}}
\caption[]{Comparison of spectra of SN 2004ab and other well studied SNe Ia at +4 d.}
\label{fig04ab_spcomp_p4}
\end{figure}

In Fig. \ref{fig04ab_spcomp_p4}, spectrum of SN 2004ab at $\sim$+4 d is compared with those of 
SN 1994D \citep{pata96}, 
SN 2003du \citep{anup05},
SN 2003cg \citep{elia06},
SN 2006X \citep{wang08},	
and
SN 2005cf \citep{gara07,wang09}
at similar epochs. The comparison spectra are obtained from SUSPECT{\footnote{\url{http://www.nhn.ou.edu/~suspect/}}} 
and WISeREP{\footnote{\url{http://www.weizmann.ac.il/astrophysics/wiserep/}}} Supernova Spectrum Archives.

Spectrum of SN 2004ab is similar to those of other SNe in comparison. 
Though, SN 2006X was a highly reddened SN similar to SN 2004ab, it had larger expansion velocity and hence most of the features are blended.
The extreme blue region of the SN 2004ab spectrum is a bit noisy. However, Ca\,{\sc ii} H\&K, Si\,{\sc ii} $\lambda$3858, Si\,{\sc ii} $\lambda$4130 features are clearly visible.  

\begin{figure}
\centering
\resizebox{\hsize}{!}{\includegraphics[trim = 0mm 80mm 0mm 0mm, clip]{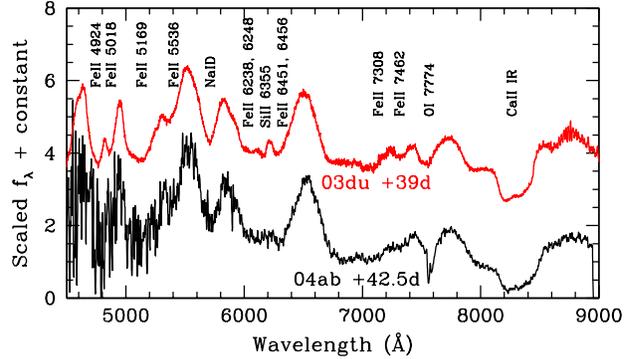}}
\caption[]{Comparison of spectra of SN 2004ab and SN 2003du at +42.5 d.}
\label{fig04ab_spcomp_p42}
\end{figure}

The spectrum of SN 2004ab at +42.5 d is compared with that of SN 2003du in Fig. \ref{fig04ab_spcomp_p42}. It is clear that spectrum of SN 2004ab is very similar to that of SN 2003du. Spectra of both the SNe are characterized by strong Fe\,{\sc ii} lines. 
Other features like Na\,{\sc i} and Ca\,{\sc ii} NIR triplet are very strong and similar in the spectra of SN 2004ab and SN 2003du. The overall spectral evolution of SN 2004ab is very similar to those of normal SNe Ia.

\subsection{Expansion Velocity}

\begin{figure}
\centering    
\resizebox{\hsize}{!}{\includegraphics{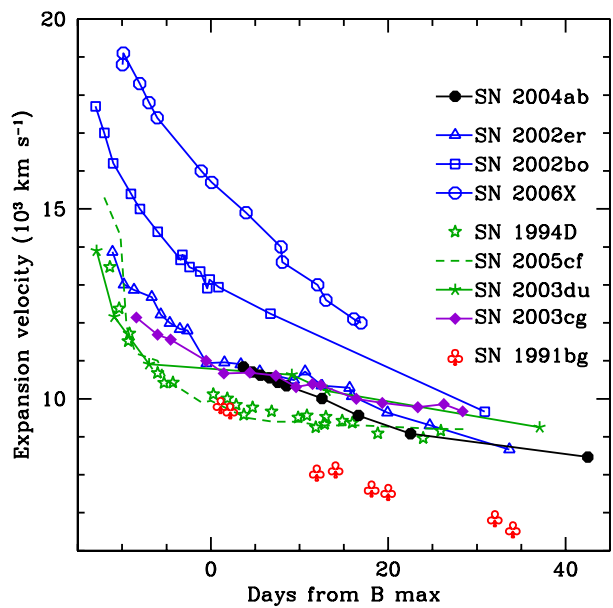}}
\caption[]{Velocity evolution of  Si\,{\sc ii} $\lambda$6355, absorption line for SN 2004ab is compared with other SNe Ia.}
\label{fig04ab_vel}
\end{figure}

Photospheric velocity  measured from Si\,{\sc ii} $\lambda$6355 absorption line in the spectra of SN 2004ab is plotted in Fig. \ref{fig04ab_vel} and compared with those of other SNe Ia. SN 2004ab was discovered late, hence, there is no early phase velocity information for this supernova. At +3.7 d, SN 2004ab has photospheric velocity of $\sim$10800 km\,s$^{-1}$, similar to SN 2002er, SN 2003cg and SN 2003du, but higher than SN 1994D, SN 2005cf  and lower than the  high velocity SNe Ia SN 2002bo and  SN 2006X. One week after $B$ band maximum, the velocity of SN 2004ab reduces  to $\sim$10500 km\,s$^{-1}$, and after two weeks it becomes $\sim$9000 km\,s$^{-1}$. At the last data point, which corresponds to +42.5 d, the velocity of Si\,{\sc ii} $\lambda$6355 absorption line is $\sim$8500 km\,s$^{-1}$.   

The  expansion velocity of normal SNe Ia shows rapid temporal evolution  during pre-maximum phase,
and after $B$ band maximum expansion velocity changes slowly. From Fig. \ref{fig04ab_vel}, it is clear that SN 2004ab follows a trend similar to that of SN 2002er. The velocity evolution of both SNe are very similar. 
  
\subsection{Spectroscopic parameter}    

\begin{figure}
\resizebox{\hsize}{!}{\includegraphics{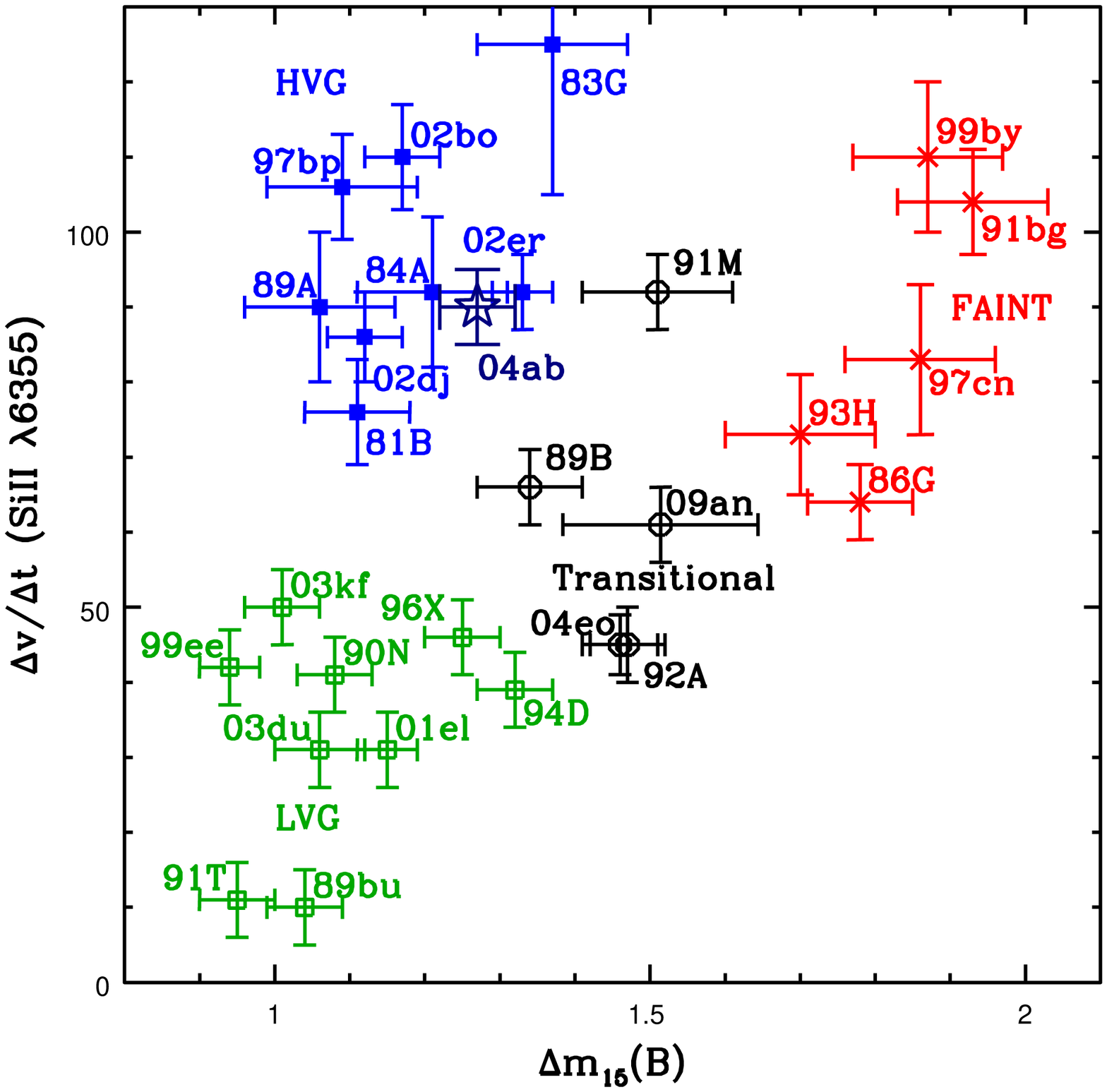}}
\caption[]{Spectroscopic subclassification of SN 2004ab based on the scheme of \citet{bene05}. Transitional events are included from  \citet{past07_eo,sahu13}.}
\label{fig04ab_vel_grad}
\end{figure}

Based on the gradient of  velocity evolution of  the Si\,{\sc ii} $\lambda$6355 line,   
\citet{bene05} grouped SNe Ia into three different subclasses. The SNe showing very slow velocity evolution and hence low velocity gradients in the post maximum phase are termed as Low Velocity Gradient (LVG; velocity gradient $\dot{v}$ $<$ 60 -- 70 km\,s$^{-1}$\,d$^{-1}$) events. SN 1994D, SN 2003du, SN 2005cf fall in this subgroup. On the other hand SNe like SN 2002er,
SN 2002bo and SN 2006X having high velocity gradients are called High Velocity Gradient (HVG) events. LVG includes
luminous and normal SNe Ia, while HVG group  has average luminosity. SN 1991bg--like under luminous SNe show a high velocity gradient, and are grouped under Faint subclass.  
 The measured velocity gradient of SN 2004ab is  90 km\,s$^{-1}$\,d$^{-1}$, which falls in the HVG subclass. The classification scheme of \citet{bene05} is shown in Fig. \ref{fig04ab_vel_grad} and the position of SN 2004ab is marked in it. SN 2004ab occupies place near SN 2002er in the clustered region of HVG. The average luminosity of SN 2004ab is also consistent with its belonging to HVG group. 

\begin{figure}
\resizebox{\hsize}{!}{\includegraphics{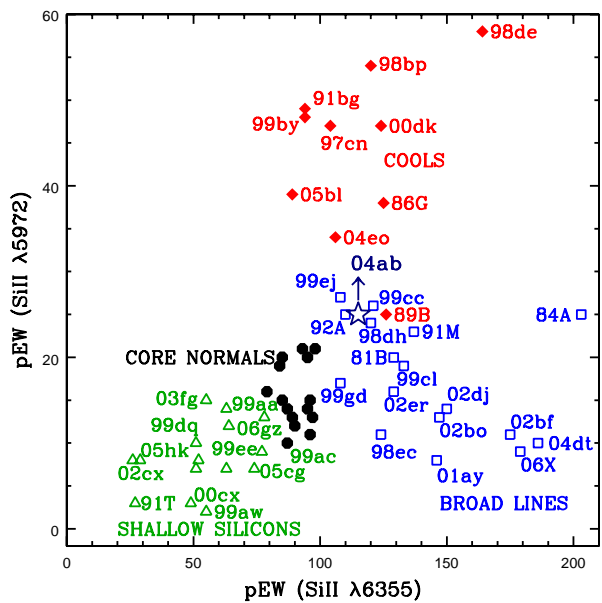}}
\caption[]{Spectroscopic subclassification of SN 2004ab based on the scheme of \citet{bran06}.}
\label{fig04ab_eqw_ratio}
\end{figure}
    
\citet{bran06} introduced another method of classifying SNe Ia spectroscopically using the  pseudo-EWs of Si\,{\sc ii} $\lambda$5972 (W(5750)) and Si\,{\sc ii} $\lambda$6355 (W(6100)) lines in the spectrum around $B$ band maximum.  In a  plot between W(5750) and W(6100) the SNe are distributed  in  four regions 
of shallow silicon (SS; SN 1991T--like), core normal (CN; SN 2003du--like), broad line (BL; SN 2002bo--like)  and cool (CL; SN 1991bg--like). The classification scheme of \citet{bran06} is shown in Fig. \ref{fig04ab_eqw_ratio} with position of SN 2004ab marked. It falls in the region occupied by BL objects. The HVG objects  have properties overlapping with the BL group. This is also applicable for SN 2004ab.

\begin{figure}
\resizebox{\hsize}{!}{\includegraphics{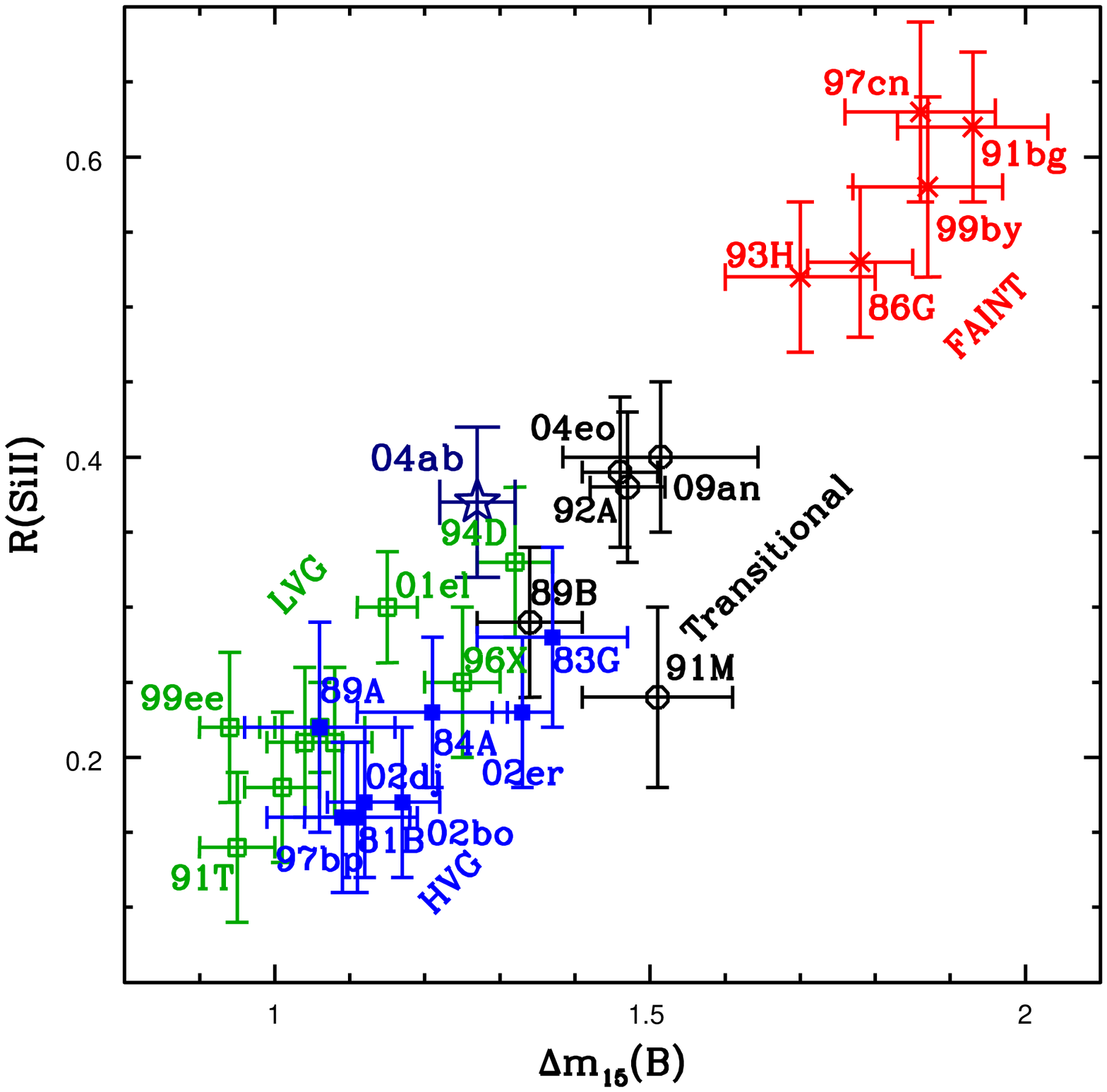}}
\caption[]{$\cal R$(Si\,{\sc ii}) ratio vs. $\Delta m_{15}(B)$ plot for SN 2004ab and other SNe Ia from \citet{bene05,past07_eo,sahu13}.}
\label{fig04ab_deltam_R}
\end{figure}

The  strength ratio of 
Si\,{\sc ii} $\lambda$5972 and Si\,{\sc ii} $\lambda$6355 lines is correlated with  the luminosity of SNe Ia \citep{nuge95,bene05}. The   value of $\cal R$(Si\,{\sc ii}) is found to be smaller for luminous objects and larger for fast declining, fainter and cooler objects. We measured the line strength ratio, $\cal R$(Si\,{\sc ii}) for SN 2004ab as 0.37 and plotted against $\Delta m_{15}(B)$ in Fig. \ref{fig04ab_deltam_R} alongwith other SNe Ia from \citet{bene05,past07_eo}; \citet*{sahu13}. SN 2004ab occupies position near SN 1994D within the diagonal strip of Fig.  \ref{fig04ab_deltam_R}.

Using the expansion velocity estimated from Si\,{\sc ii} line in the spectrum close to maximum light, \citet{wang09hv} grouped SNe Ia into two classes: those having high velocity (HV), $v$ $\geq$ 
11800 km\,s$^{-1}$ and others having normal velocity (NV), with an average of $\langle v \rangle$ = 10600 km\,s$^{-1}$. Since, SN 2004ab was discovered late, our first spectrum was obtained 3.7 d after maximum light. Hence, exact value of velocity at maximum light could not be determined. However, from the observed trend in velocity evolution (refer Fig. \ref{fig04ab_vel}), it is inferred that at maximum light it would have a velocity of $\sim$11000 km\,s$^{-1}$, hence can be considered as NV type. \citet{wang09hv} found that $R_V$ is lower ($\sim$1.6)  for HV  as compared to NV ($\sim$2.4) subgroup in  their sample. However, \citet{fole11} found that after  
excluding highly reddened SNe having $(E(B-V) > 0.35)$ from the sample of \citet{wang09hv}, both NV and HV subgroups are consistent with same value of $R_V$ = 2.5. SN 2004ab seems to be a normal velocity SN and shows very low value of $R_V$ = 1.45.

Velocity evolution and gradient of SN 2004ab matches well with that of SN 2002er (refer Fig. \ref{fig04ab_vel} and \ref{fig04ab_vel_grad}). Both  SNe show high gradient in their velocities. However, there is no similarity in the line strength ratio, $\cal R$(Si\,{\sc ii}) of these two SNe. The $\cal R$(Si\,{\sc ii}) of SN 2004ab is higher than that of SN 2002er. The $\cal R$(Si\,{\sc ii}) value of SN 2004ab is more like LVG SNe, while SN 2002er follows other HVG SNe.
This is because, the Si\,{\sc ii} $\lambda$5972 line is stronger in SN 2004ab compared to SN 2002er. Many HVG SNe overlap with the BL type and this is also true for SN 2004ab. But unlike most other BLs, which  have strong Si\,{\sc ii} $\lambda$6355 line and relatively weaker Si\,{\sc ii} $\lambda$5972,  SN 2004ab has stronger Si\,{\sc ii} $\lambda$5972 line and weaker Si\,{\sc ii} $\lambda$6355 line.

\section{Summary}
\label{sec04ab_summary} 

Optical photometric and spectroscopic analyses of SN 2004ab are presented. 
SN 2004ab is a highly reddened normal type Ia supernova with $E(B - V )_\text{total}$ = 1.70 mag. 
The intrinsic decline rate parameter of SN 2004ab is $\Delta m_{15}(B)_\text{true}$ = 1.27.
The photospheric velocity evolution measured from Si\,{\sc ii} $\lambda$6355, absorption line is similar to that of SN 2002er. The Si\,{\sc ii} $\lambda$6355 velocity gradient is estimated as $\dot{v}$ = 90 km\,s$^{-1}$\,d$^{-1}$ indicating that SN 2004ab is a member of HVG subgroup.  The pseudo-EWs of Si\,{\sc ii} $\lambda$5972 and $\lambda$6355 absorption lines suggest that SN 2004ab is a BL type.
The line strength ratio  $\cal R$(Si\,{\sc ii}) is  0.37,  higher than  those of other BLs having similar $\Delta m_{15}(B)$. This is due to higher strength of Si\,{\sc ii} $\lambda$5972 in SN 2004ab.
Using CCM model, the ratio of total-to-selective extinction for host galaxy NGC 5054, in the direction of SN 2004ab,  is derived as $R_V$ = 1.41, which is much lower than that of Milky Way. The derived extinction is also consistent with power law extinction model $A_\lambda/A_V$ = ($\lambda/\lambda_V)^p$ with $p \sim -2.2$. SN 2004ab peaked at an absolute magnitude of $M_{B}^{\text{max}}$ = $-$19.31 $\pm$ 0.25 mag.
 Peak bolometric luminosity of $\log L_\text{bol}^\text{max}$ = 43.10 $\pm$ 0.07 erg\,s$^{-1}$ suggests that 0.53 $\pm$ 0.08 M$_\odot$ of $^{56}$Ni was synthesized in this explosion.  Though, SN 2004ab is a highly reddened supernova, its absolute luminosity,  after correcting for  extinction using  non standard extinction law,  is similar to a normal SN Ia and follows the empirical {\it Luminosity decline rate relation}.  

\section*{Acknowledgements}
We thank the anonymous referee for going through the draft carefully and providing  constructive suggestions.
NKC is thankful to the Director and Dean of IIA, Bangalore for local hospitality and facilities provided. We are thankful to Ramya S. and Jessy J. for  their assistance during the observations and to Shubham Srivastav for help in fitting the light curves using SNooPy code. All the observers of the 2-m HCT (IAO-IIA), who kindly provided part of their observing time for supernova observations, are  thankfully acknowledged. This work has made use of the NASA/IPAC Extragalactic Database (NED) which is operated by Jet Propulsion Laboratory, California Institute of Technology, under contract with the National Aeronautics and Space Administration. We have also made use of the Lyon-Meudon Extragalactic Database (LEDA), supplied by the LEDA team at the Centre de Recherche Astronomique de Lyon, Observatoire de Lyon. We acknowledge use of the
SNooPy software package developed by the Carnegie Supernova Project (CSP) and the Online Supernova Spectrum Archive (SUSPECT) initiated and maintained at the Homer L. Dodge Department of Physics and Astronomy, University of Oklahoma.

\bibliographystyle{mn2e_warrick}
\bibliography{references.bib}

\label{lastpage}
\end{document}